\documentclass[%
 reprint,
 amsmath,amssymb,
 aps, 
]{revtex4-2}
\usepackage{xcolor}
\usepackage{graphicx}
\usepackage{dcolumn}
\usepackage{bm}

\begin{document}

\preprint{APS/123-QED}

\title{Stability and Spatial Autocorrelations of Suspensions of Microswimmers with Heterogeneous Spin}

\author{Yonatan Ashenafi}
 \email{yashenaf@ualberta.ca}
 \affiliation{
        Department of Mathematical and Statistical Sciences\\
        University of Alberta\\
        Edmonton, T6G 2R3 AB, Canada
        }   

\date{\today}

\begin{abstract}
Hydrodynamical interactions of active micro-particles are pervasive in our planet's fluid environments. Hence, understanding the interactions of these self-propelled particles is essential for science and engineering. In this paper the suspensions of active swimmer-rotor particles have been mathematically modeled by extending a previously developed stochastic kinetic theory to analyze heterogeneous collections of microswimmers and microrotors with multiple spin velocity populations. The paper uses this modeling approach to derive insights on large scale properties such as suspension instabilities and spatial correlations of the active particles and highlights the role of active particle rotations on the behavior of the suspension. {This paper will focus on analytically explaining emergence of pattern formation observed for self-propelled micro-particles using numerical and lab experiments.}
\end{abstract}

\maketitle


\section{\label{sec:level1}Introduction}

When active microswimmers are close enough to each other and their suspensions are concentrated, their dynamics become interdependent through the fluid medium and they interact. Furthermore, active
particles are known to have collective dynamics beyond the individual particles \cite{Dunkel,Sokolov} and to achieve enhanced mixing \cite{Breuer,Mino} through these interactions. Hence, biological fluid mechanics requires understanding of {these} hydrodynamical interactions.  Understanding the interactions of the active particles analytically is challenging because the systems of differential equations governing each particle become intractably coupled and highly non-linear. Simulations of individual particle's dynamics are also demanding since the number of particles being considered makes the computational cost high. However this challenge has been overcome in many recent works. For example, results that show interesting phenomena such as large scale coherent motion for high concentrations of swimmers have been discovered \cite{Hernandez-simulation, Underhill-simulation}. 

The dynamic suspensions of active particles have been mathematically modeled and computationally studied extensively \cite{Sokolov, Roper2013, Dean1996, Koch}. {Furthermore the dynamic process of interacting polymers and filaments have also been investigated with simulations \citep{Winkler}}. Continuum models that overcome the computational cost of individual particle's dynamics simulation and give insight into large scale properties such as suspension instabilities. The models were often described by mean field equations \cite{Saintillan-Shelley,Simhastress}. The mean field equation are deterministic by virtue of assuming the suspensions have infinite number of particles and corresponding statistical laws. This assumption is clearly inaccurate and leads to ignoring the real noise {and correlations} that the system exhibits \citep{Morozov2}. More recently, stochastic field equations that take into account the fluctuations of a finite system of particles have been developed giving more insights about spatial autocorrelations of the system's variables and effects on inert particles \cite{Qian}.

In this paper we continue this development for a broader class of active particles with swimmer-rotors being our primary targets. The study of the collective behavior of swimmer-rotors has been extensively undertaken for various circumstances such as local alignment interactions, and phase separation for temporally biphasic particles \cite{Cates2015, Liebchen2017, Vicsek1995}. We will also focus on heterogeneous collections paying particular attention to angular velocities. The heterogeneity in angular velocity is readily observed in nature. For example we see it with colonial algae swimmers with varying degrees of collective torque \cite{Kirkgaard} and with single celled swimmers with stimulus driven motion or taxis \cite{pedley_kessler}. 

In terms of order we will first construct the equations of motion of individual active particles in section \ref{Model}. Then, we will look into how a stochastic kinetic equation is derived from the mesoscopic swimmer SDEs and fluid flow equations following previous approaches in the same section \cite{Qian}. Next, having linearized our stochastic kinetic equation about a uniform isotropic state, we study its stability as we perturb important parameters in section \ref{Stability Analysis}. Finally, we will discuss interesting spatial auto-correlations that the swimmers develop and how {these} correlations have physical significance in section \ref{Correlations} and conclude with a discussion in section \ref{discussion}.

\subsection{Mathematical Model} \label{Model}

\begin{figure}[h!]
\begin{center}
    \includegraphics[height=40mm,width=80mm]{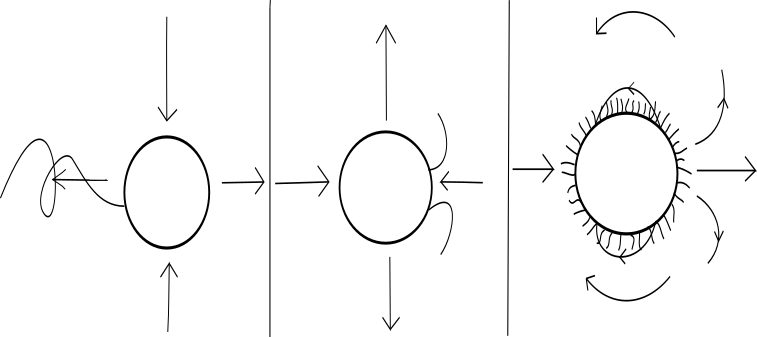}
\caption{Velocity field around swimmers moving rightward. Left to right: Pusher, puller, neutral squirmer.}\par
\label{pusherpullersquirmer}
\end{center}
\end{figure}

Microswimmers are known to have a wide variety of swimming mechanisms. These {mechanisms classify swimmers} into three catagories: pullers, pushers,and neutral squirmers. {They are categorized by the fluid velocity field they generate around them. The velocity field differences are shown in figure \ref{pusherpullersquirmer}.}
The difference between these mechanisms shows up in our model in the forcing of the Stokes' fluid flow equations. In our modeling we will work on a general framework that covers pushers and pullers but we will focus on pushers for our results. 

The hydrodynamic interactions of active particles has been studied extensively in the past. One recent development was the framework given in \cite{Qian}. We adopt this framework and start with active microswimmers'  position ($\boldsymbol{x}_m$) and orientation ($\theta_m$). We then append the model by constructing the swimmers with an active torque which, when balanced with rotational drag, leads to an active component for the angular velocity.\\
Including the active angular velocity component($\hat{\mu}$) we have:
\begin{equation} \label{eq_c}
    \begin{aligned}
    d\boldsymbol{x}_m(\hat{t})=&(\mathbf{\hat{u}_m}(\hat{t})+\hat{\nu}_m(\cos(\theta_m),\sin(\theta_m)))d\hat{t}\\&
    +\sqrt{2\hat{D}_t}d\mathbf{W_{x,m,\hat{t}}}
    \end{aligned}
\end{equation}
\begin{equation} \label{eq_d}
    \begin{aligned}
    d\theta_m(\hat{t})=(\hat{\omega}_m(\hat{t})+\hat{\mu}_m )d\hat{t}+\sqrt{2\hat{D}_r}d\mathbf{W_{\theta,m,\hat{t}}}
    \end{aligned}
\end{equation}

$\hat{\nu}_m$ and $\hat{\mu}_m$ are the translational and rotational active speed respectively of particle m. $\mathbf{\hat{u}_m}$ and $\mathbf{\hat{\omega}_m}$ is the velocity and vorticity respectively on particle m caused by the ambient stokes flow. $\hat{D}_{t,m}$ and $\hat{D}_{r,m}$ are the translational and rotational diffusion coefficients respectively. There we see that the non-thermal noise that is due to the molecular motors of the flagella plays the dominant role for the diffusion. 

{Note that the model is two-dimensional in space with an additional orientation dimension for the rotations. These was picked to avoid the complications of rotational torque along multiple axis of rotation.   Furthermore, it is not entirely unrealistic since some motile protozoa such as raphid diatoms are known to swim and feed on top of biofilm substrates which can be modeled to be piecewise planar \citep{IWASA1972552}.}

An assumption that was made in \cite{Qian} and will be made here is that the swimming speed and the diffusion (both translational and angular) are constant across different swimmers.  This can be relaxed as needed in the framework that follows. As a starting point we will remove only the assumption on $\hat{\mu}_m$ and proceed. In this case we have:
    \begin{equation} 
    \begin{aligned}
    \hat{\nu}_m=\hat{\nu}, \hspace{0.3 cm} \hat{D}_{t,m}=\hat{D}_{t},\hspace{0.3 cm}  \hat{D}_{r,m}=\hat{D}_{r}
    \end{aligned}
\end{equation}

Next, we non-dimensionalize and scale the system with:\\
1) ${t}=\frac{\hat{t}}{t_c}$ where $t_c=\frac{1}{\hat{D}_r}$\\
2) $x=\frac{\tilde{x}}{x_c}$ where $x_c=\sqrt{\frac{L^2}{N_p}}$\\
Where $L$ is the size of the system and $N_p$ is the number of active particles. The space scaling focuses the equations on the length scale of a swimmer's box in a uniformly spaced suspension while the time scaling focuses the equations on the time scale of the duration taken to reach orientational equilibrium (isotropy). 
We now have: 
\begin{equation} \label{nondim-eq_c}
    \begin{aligned}
    d\boldsymbol{x}_m({t})=&(\mathbf{u_m}({t})+\nu(\cos(\theta_m),\sin(\theta_m)))d{t}\\&
    +\sqrt{2{D}_t}d\mathbf{W_{x,m,t}}
    \end{aligned}
\end{equation}
\begin{equation} \label{nondim-eq_d}
    \begin{aligned}
    d\theta_m({t})=(\omega_m({t})+\mu_m )dt+\sqrt{2}d\mathbf{W_{\theta,m,{t}}}
    \end{aligned}
\end{equation}
Where $\mathbf{u_m}=\frac{t_c}{x_c}\mathbf{\hat{u}_m}$, $\nu=\frac{t_c}{x_c}\hat{\nu}$, $\omega_m=t_c\hat{\omega}_m$, $\mu_m=t_c\hat{\mu}_m$, ${D}_t=\frac{t_c}{x_c^2}\hat{D}_t$.

Next we look at the suspension microhydrodynamics. The flow created by a suspension of mircoswimmers creates hydrodynamic interactions and alters the motion of passive tracers.
\cite{Saintillan} used mean field theory based on point-dipole models to explain this.

The particle density of the swimmers (whose locations and orientations are prescribed in dimensional terms by $\mathbf{\tilde{x}_m}$ and $\theta_m$) at position $\tilde{x}$ with a given orientation $\theta$ is given by:
    \begin{equation}
    \hat{\psi}(\boldsymbol{\tilde{x}},\theta,\tilde{t})=\sum\limits_{m}\delta(\mathbf{\tilde{x}}-\tilde{\boldsymbol{x}}_m(\tilde{t}))\delta(\theta-\theta_m(\tilde{t}))
    \end{equation}
For a finite number of swimmers, the Fourier transform of the particle density($\hat{\psi}$) is:
    \begin{equation}
    \hat{\psi}_{\mathbf{k},l}(t)=\frac{1}{2\pi L^2}\sum\limits_{m}\exp(-i\mathbf{k}.\mathbf{\tilde{x}_m(\tilde{t})}-il\theta_m(\tilde{t}))
    \end{equation}
From here on we will refer to the non-dimensional time as $t$ and not $\tilde{t}$. Using Ito's lemma and inverting the Fourier transform (as has been done by others in the past) we find the dimensional stochastic kinetic equation for the phase space density of the swimmers. 

We then non-dimensionalize the equation. First, we nondimensionalize the independent variables. Second, we nondimensionalize the phase space density $\hat{\psi}$ with the scaling factor $\psi_c=\frac{N_p}{2\pi L^2}$ where $N_p$ is the total number of swimmers and L is the length of the square containing the swimmers. We then have $\psi=\frac{\hat{\psi}}{\frac{N_p}{2\pi L^2}}$ as the dimensionless phase space density. The dimensionless phase space density equation is:
\begin{equation} \label{SPDE}
\begin{aligned}
\frac{\partial {\psi}(\boldsymbol{x},\textbf{n},t)}{\partial t}=&-\nabla_{\boldsymbol{x}}\cdot(\dot{\boldsymbol{x}}{\psi}) -\nabla_{\textbf{n}}\cdot(\dot{\textbf{n}}{\psi})+D_t\nabla^2_{x}{\psi}+\nabla_n^2{\psi}\\&
+\Xi(\boldsymbol{x},\textbf{n},t)
\end{aligned}
\end{equation} 
\begin{equation} \label{SPDEb}
\begin{aligned}
\Xi dt=\mathbf{\nabla}\cdot[\sqrt{2D_t{\psi}}d\textbf{U}(\boldsymbol{x},\theta,t)]+\partial_\theta[\sqrt{2{\psi}}d\textbf{V}(\boldsymbol{x},\theta,t)]
\end{aligned}
\end{equation} 
Where
\begin{equation} \label{SPDE_extra}
\begin{aligned}
\langle d\textbf{U}(\boldsymbol{{x}},\theta, t) d\textbf{U}(&\boldsymbol{{x'}},\theta', t)\rangle=\\&\begin{pmatrix} 1 & 0 \\ 0 & 1 \end{pmatrix}
\cdot \delta(\boldsymbol{{x-x'}},\theta-\theta',t-t')dtdt'
\end{aligned}
\end{equation}
\begin{equation} \label{SPDE_extra2}
\begin{aligned}
\langle dV(\boldsymbol{{x}},\theta, t)dV(\boldsymbol{{x'}},\theta', t) \rangle=\delta(\boldsymbol{x-x'},\theta-\theta',t-t')dtdt'
\end{aligned}
\end{equation} 
The translational velocity is $\mathbf{\dot{x}}=\nu\boldsymbol{{n}}+\mathbf{u}$ and the angular velocity is given by:
\begin{equation} \label{ang vel}
\begin{aligned}
\boldsymbol{\dot{n}}=\mathbf{(I-\boldsymbol{{n}}\boldsymbol{{n}})\cdot(\gamma E-W)\cdot \boldsymbol{{n}}}+\begin{pmatrix}
-n_2 \\
n_1
\end{pmatrix}{\mu}
\end{aligned}
\end{equation} 
  
 $\mu$ is a fixed quantity signifying a single active angular velocity for each microswimmer. For the general case where each swimmer can have a different active angular velocity the active term in (\ref{ang vel}) becomes convolved with $\hat{\psi}$. 
 
Next, we define the Stokes' equation including the flow induced stresslet. 
\begin{equation} \label{stokes}
\begin{aligned}
&-\eta \nabla^2_{x}\hat{\mathbf{u}}+\nabla_{x}q=\nabla_{x}\cdot\Sigma\\&
\Sigma=d\int \hat{\psi}(\boldsymbol{{nn}}-I/2)d\boldsymbol{{n}}+\hat{\Sigma}_{flow}
\end{aligned}
\end{equation} 
We non-dimensionalize the Stokes' equation to get: 
\begin{equation} \label{stokes2}
\begin{aligned}
&-\nabla^2_{x}\mathbf{u}+\nabla_{x}q=\nabla_{x}\cdot \Big[\frac{p}{2\pi}\int {\psi}(\mathbf{nn}-I/2)d\mathbf{n}+\Sigma_{flow}\Big]
\end{aligned}
\end{equation} 
Here $p=-\frac{dN_p}{\eta \hat{D_r} L^2}$. p is a dimensionless parameter that is negative for pushers and positive for pullers. It is a product of effective strain on the fluid due to a microswimmer, time span for reaching angular equilibrium, and overall number density. We notice that $p$ can be small depending on the system's number density, the fluid's viscosity, and the particle's rotational diffusion and stresslet coefficient. 

The first term in $\Sigma$ is the active stress and the later forms the flow-induced stress. $p>0$ is the stress-let strength. The active stress is found by following past procedures to identify the force dipole of each swimmer. Namely, we start with finding the centroid $x_f$ at which the resultant forcing $F^{(0)}_{m}=F^{(0)}$ (assumed constant across particles) is located. Doing this we get: 
\begin{equation} 
\begin{aligned}
x_f=\frac{a}{F^{(0)}}\sum\limits_{i=1}^{N}F_{i}^{(0)}\begin{pmatrix}\cos(2\pi[\frac{i-\frac{1}{2}+\frac{s_{i,m}}{l}}{N}]+\theta_{i,m}^0) \\ \sin(2\pi[\frac{i-\frac{1}{2}+\frac{s_{i,m}}{l}}{N}]+\theta_{i,m}^0)\end{pmatrix}
\end{aligned}
\end{equation} 
We now get the corresponding drag force's location. We can take a simple approach of assuming the drag force is distributed in a manner that sharply and evenly vanishes across the swimmer surface (like a regularized Dirac delta function for instance) starting from the point that is along the forcing vector and use a distributed load integration approach to get the surface point that is along the forcing to be the approximate location of the drag. 


We now have a dipole representation for the swimmers. We then use a previously developed approach to get the stress \cite{Simhastress}. 

$\Sigma_{flow}$ is the flow induced stresslet. With a dilute suspension assumption we assume $(\Sigma_{flow})_{i,j}$ to be negligible.

The next step is linearization. To study the stability of the uniform isotropic suspension and the emergence of correlations from such a suspension, we can expand ${\psi}$ ($\psi=1+\epsilon\phi+O(\epsilon^2)$) and linearize the SPDE like is done on \cite{Qian}. We then Fourier expand and see that the matrix for the modes ($L_{\mathbf{k},l,l'}$) is modified on its diagonal due to $\hat{\mu}$ being non-zero. Namely we would get:
\begin{equation} \label{matrix}
\begin{aligned}
L_{\mathbf{k},l,l'}= & -\nu\frac{i}{2}ke^{i\theta_{k}}\delta_{l,l'-1}- \nu\frac{i}{2}ke^{-i\theta_{k}}\delta_{l,l'+1}-D_{t}k^2\delta_{l,l'}\\&+[-l^2-{il\mu}]\delta_{l,l'}
+\delta_{l,2}\frac{p{\gamma}}{8}(-\delta_{l',2}+e^{-i4\theta_{k}}\delta_{l',-2})\\&
+\delta_{l,-2}\frac{p{\gamma}}{8}(\delta_{l',2}e^{i4\theta_{k}}-\delta_{l',-2})
\end{aligned}
\end{equation} 
 Next we look at two populations of swimmers with different angular velocities. Namely with angular velocities $\hat{\mu}_1$ and $\hat{\mu}_2$. After reconsidering the kinetic equation and non-dimensionalizing we get the following system:
\begin{equation} \label{SPDE2}
\begin{aligned}
\frac{\partial \psi_{\bar{i}}(\boldsymbol{x},\textbf{n},t)}{\partial t}=&-\nabla_{\boldsymbol{x}}\cdot(\dot{\boldsymbol{x}}\psi_{\bar{i}})
-\nabla_{\textbf{n}}\cdot(\dot{\textbf{n}}_{\bar{i}}\psi_{\bar{i}})+D_t\nabla^2_{x}\psi_{\bar{i}}\\&
+D_r\nabla_n^2\psi_{\bar{i}}+\Xi_{\bar{i}}(\boldsymbol{x},\textbf{n},t)
\end{aligned}
\end{equation} 
\begin{equation} \label{SPDE2b}
\begin{aligned}
\Xi_i dt=\mathbf{\nabla}\cdot[\sqrt{2D_t\psi_i}d\textbf{U}(\boldsymbol{x},\theta,t)]+\partial_\theta[\sqrt{2{\psi}_i}d\textbf{V}(\boldsymbol{x},\theta,t)]
\end{aligned}
\end{equation} 
Where $d\textbf{U}$ and $dV$ can be given with (\ref{SPDE_extra}). 

For the Stokes' flow we get:
\begin{equation} \label{stokes3}
\begin{aligned}
&-\nabla^2_{x}\mathbf{u}+\nabla_{x}q=\nabla_{x}\cdot \Big[\frac{p}{2\pi}\int (\sum\limits_{i=1}^2{\psi}_i)(\mathbf{nn}-I/2)d\mathbf{n}\Big]
\end{aligned}
\end{equation} 
Once we Fourier expand the kinetic equation and linearize it we get an inhomogeneous system for each sub population with the inhomogenity coming from the other sub population(s). The deterministic part of the system is:

\begin{equation}  \label{system 1}
\begin{aligned}
\frac{\partial \mathbf{{\psi}_{\bar{i},\mathbf{k}}}}{\partial t}=
\mathbf{L_{\bar{i},\mathbf{k}}}\mathbf{{\psi}_{\bar{i},\mathbf{k}}}+\sum\limits_{\bar{j}\neq \bar{i}} \mathbf{M_{\bar{j},\mathbf{k}}}\mathbf{{\psi}_{\bar{j},\mathbf{k}}}
\end{aligned}
\end{equation} 
Where

\begin{equation} \label{System 2}
\begin{aligned}
L_{\bar{i},\mathbf{k},l,l'}&=  -\nu\frac{i}{2}ke^{i\theta_{k}}\delta_{l,l'-1}-\nu \frac{i}{2}ke^{-i\theta_{k}}\delta_{l,l'+1}-D_{t}k^2\delta_{l,l'}\\&
+[-l^2-{il\mu_{\bar{i}}}]\delta_{l,l'}+\delta_{l,2}\frac{p{\gamma}}{8}(-\delta_{l',2}+e^{-i4\theta_{k}}\delta_{l',-2})\\&
+\delta_{l,-2}\frac{p{\gamma}}{8}(\delta_{l',2}e^{i4\theta_{k}}-\delta_{l',-2})
\end{aligned}
\end{equation} 
\begin{equation} \label{System 2b}
\begin{aligned}
M_{\bar{j},\mathbf{k},l,l'}&=\delta_{l,2}\frac{p\gamma}{8}(-\delta_{l',2}+e^{-i4\theta_{k}}\delta_{l',-2})\\&
+ \delta_{l,-2}\frac{p{\gamma}}{8}(\delta_{l',2}e^{i4\theta_{k}}-\delta_{l',-2})
\end{aligned}
\end{equation} 

\section{Stability Analysis}  \label{Stability Analysis}
Now that we have the system of equations for the suspension we consider the parameter regimes where perturbations of uniform isotropic suspensions are stable. We will do is for a few interesting cases. 

\subsection{Active Particles with small $p\gamma$} \label{Nearly Circular Stability}
{Nearly circular elliptical swimmers with low eccentricity are the two dimensional analogs of nearly spherical particles. Nearly spherical microswimmer-rotors are found abundantly in nature with examples such as Volox, and colonial algae such as Choanoflagellates\cite{Kirkgaard,Bonner}. Particles that are shaped nearly circularly suspended in a dilute suspension correspond to $p\gamma \ll 1$ and we will use this regime next to study suspension stability.}

We will first work out the leading order stability situation. When we have a circular shape ($\gamma=0$) as in our model we will get a decoupling of the populations of different angular velocities. Moreover, the linearized kinetic equations will be independent of the Stokes' flow and we get the following tridiagonal matrix for a given angular velocity population:
\begin{equation} \label{matrix3}
\begin{aligned}
L= & -\nu\frac{i}{2}ke^{i\theta_{k}}\delta_{l,l'-1}-\nu \frac{i}{2}ke^{-i\theta_{k}}\delta_{l,l'+1}-D_{t}\mathbf{k}^2\delta_{l,l'}\\&
+[-l^2-{il\mu_i}]\delta_{l,l'}
\end{aligned}
\end{equation} 
Instead of looking for the eigenvalues of this matrix directly we inverse Fourier transform back for angular orientation(keeping the spatial modes as they are) and we get the PDE:
\begin{equation}
\begin{aligned}
    \frac{\partial \Psi}{\partial t}=&-\nu(\cos(\theta),\sin(\theta))^T(i\mathbf{k}\Psi)- {D}_{t}(k_{1}^2+k_{2}^2)\Psi\\&
    + \Psi_{\theta \theta}-\hat{\mu_i}\Psi_{\theta}
\end{aligned}
\end{equation}
Note that the above equation is using the independent variable $\theta$ whereas previously we were using $\mathbf{n}$. Using separation of variables we get the following eigenvalue problem for orientation.
\begin{equation}
\begin{aligned}
    &f''(\theta)-\big(\mu_i\big)f'(\theta)-(\lambda+\nu(\cos(\theta),\sin(\theta))^T(i\mathbf{k})+\\& {D}_{t}(k_{1}^2+k_{2}^2))f(\theta)=0\\&
    f(\pi)=f(-\pi), \hspace{0.6cm} f'(\pi)=f'(-\pi)
\end{aligned}
\end{equation}
 
We will get the following general solution to the ODE:  
\begin{equation}
\begin{aligned}
    & e^{\frac{\mu\theta}{2}}C_1\text{ce}_{n}(\frac{\arctan[\frac{k_2}{k_1}]-\theta}{2} , 2i\sqrt{k_1^2+k_2^2})\\& +  e^{\frac{\mu\theta}{2}}C_2\text{se}_{n}(   \frac{\arctan[\frac{k_2}{k_1}]-\theta}{2},2i \sqrt{k_1^2+k_2^2})
\end{aligned}
\end{equation}
where $[-({4\lambda_{n}+4D_t(k_1^2+k_2^2)})-{\mu_i^2}]$ must equal the countably infinite characteristic numbers $a_n$ and $b_n$ for the cosine elliptic and sine elliptic functions respectively. Applying the periodic BC we get the equations:
\begin{equation} \label{ef1}
\begin{aligned}
    (i) & \hspace{0.5 cm}  C_1 \text{ce}_{n}(\frac{\arctan[\frac{k_2}{k_1}]-\pi}{2} ,2i{|\mathbf{k}|})\\&
    +  C_2\text{se}_{n}(\frac{\arctan[\frac{k_2}{k_1}]-\pi}{2} ,2i{|\mathbf{k}| })\\&
    =C_1 \text{ce}_{n}(\frac{\arctan[\frac{k_2}{k_1}]+\pi}{2} ,2i {|\mathbf{k}| })\\&+  C_2 \text{se}_{n}(\frac{\arctan[\frac{k_2}{k_1}]+\pi}{2} ,2i{|\mathbf{k}| })
\end{aligned}
\end{equation}
\begin{equation} \label{ef2}
\begin{aligned}
    (ii) & \hspace{0.5 cm} C_1(\frac{d(\text{ce})}{d\theta})_{n}(\frac{\arctan[\frac{k_2}{k_1}]-\pi}{2} ,2i{|\mathbf{k}|})\\&
    +  C_2(\frac{d(\text{se})}{d\theta})_{n}(\frac{\arctan[\frac{k_2}{k_1}]-\pi}{2} ,2i{|\mathbf{k}|})\\&
    =C_1(\frac{d(\text{ce})}{d\theta})_{n}(\frac{\arctan[\frac{k_2}{k_1}]+\pi}{2} ,2i{|\mathbf{k}|})\\&
    +  C_2(\frac{d(\text{se})}{d\theta})_{n}(\frac{\arctan[\frac{k_2}{k_1}]+\pi}{2} ,2i{|\mathbf{k}|})
\end{aligned}
\end{equation}

The characteristic number equations show us that the translational diffusion and the squared angular velocities  stabilize the system. Specifying the angular velocities, using (\ref{ef1},\ref{ef2}) for each n, and enforcing that $C_1$ and $C_2$ are nontrivial we find $\mathbf{k}$ for each n. We then use the definition of the characteristic numbers to find the eigenvalues ($\lambda_n$). We can then find the corresponding eigenfunctions. However, this is not an objective for this paper. 


\subsection{Slow Swimming Active Particles}  \label{Slow Swimming Stability}
{Besides nearly circular mircoswimmers we are interested in slow swimming active particles since they present an interesting comparison with numerical simulations done by \cite{Yeo2015}. In that paper they use a force coupling numerical solver to look at the suspension dynamics micro-rotors with no swimming. The method includes far field multi-body interactions with regularized low-order multipoles and near field lubrication interactions with analytical solutions. Generally speaking, they found that same spin swimmers cluster together while opposite spin swimmers segregate. We endeavored to show this effect using our analytical method.}

\begin{figure}[h!]	 
\includegraphics[height=55mm,width=\linewidth]{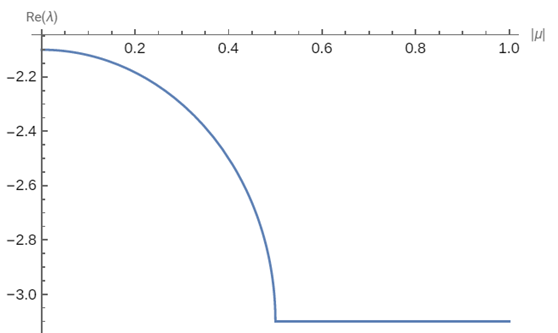}
\par
\includegraphics[height=55mm,width=\linewidth]{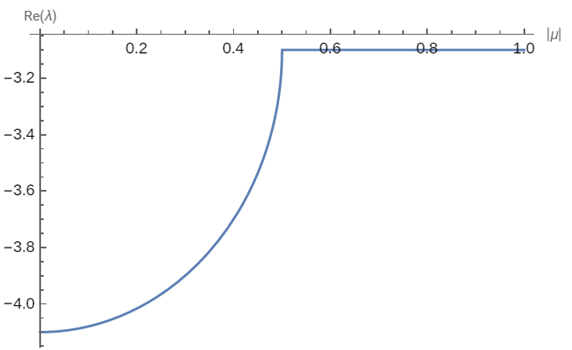}\par
\caption{Real part of the two non-trivial eigenvalues from the system given by (\ref{stabsystemslow1}-\ref{stabsystemslow4}) with ${D}_tk^2=0.1$ and $\frac{p\gamma}{8}=-0.5$ \cite{WolframProgrammingLab}.}
\label{two eigenvalues}
\end{figure}

First we will work out the leading order case with no swimming. Here we can solve the matrix system and will do so. Using the system defined on (\ref{system 1}) and (\ref{System 2}) as a reference, the system's matrix will be given by:
\begin{equation} 
\begin{aligned}
L_{\bar{i},\mathbf{k},l,l'}= & -{D_{t}}k^2\delta_{l,l'}
+[-l^2-{il{\mu}_{\bar{i}}}]\delta_{l,l'}\\&+\delta_{l,2}\frac{p{\gamma}}{8}(-\delta_{l',2}+e^{-i4\theta_{k}}\delta_{l',-2})\\&
+\delta_{l,-2}\frac{p{\gamma}}{8}(\delta_{l',2}e^{i4\theta_{k}}-\delta_{l',-2})
\end{aligned}
\end{equation} 
\begin{equation} 
\begin{aligned}
M_{\bar{j},\mathbf{k},l,l'}=&\delta_{l,2}\frac{p\gamma}{8}(-\delta_{l',2}+e^{-i4\theta_{k}}\delta_{l',-2})\\&
+ \delta_{l,-2}\frac{p{\gamma}}{8}(\delta_{l',2}e^{i4\theta_{k}}-\delta_{l',-2})
\end{aligned}
\end{equation}
The angular modes at $l=-2$ and $l=2$ for the two spin populations will have a reduced system given by:
\begin{equation} \label{stabsystemslow1}
\begin{aligned}
&\frac{d\psi_{1,\mathbf{k},-2}}{dt}=(-{D}_tk^2-4+2i{\mu}_{1}-\frac{p\gamma}{8})\psi_{1,\mathbf{k},-2}\\&
+\frac{p\gamma}{8}\psi_{1,\mathbf{k},2}-\frac{p\gamma}{8}\psi_{2,\mathbf{k},-2}+\frac{p\gamma}{8}\psi_{2,\mathbf{k},2}
\end{aligned}
\end{equation} 
\begin{equation} \label{stabsystemslow2}
\begin{aligned}
&\frac{d\psi_{1,\mathbf{k},2}}{dt}=(-{D}_tk^2-4-2i{\mu}_{1}-\frac{p\gamma}{8})\psi_{1,\mathbf{k},2}\\&
+\frac{p\gamma}{8}\psi_{1,\mathbf{k},-2}+\frac{p\gamma}{8}\psi_{2,\mathbf{k},-2}-\frac{p\gamma}{8}\psi_{2,\mathbf{k},2}
\end{aligned}
\end{equation} 
\begin{equation} \label{stabsystemslow3}
\begin{aligned}
&\frac{d\psi_{2,\mathbf{k},-2}}{dt}=(-{D}_tk^2-4+2i{\mu}_{2}-\frac{p\gamma}{8})\psi_{2,\mathbf{k},-2}\\&
+\frac{p\gamma}{8}\psi_{2,\mathbf{k},2}-\frac{p\gamma}{8}\psi_{1,\mathbf{k},-2}+\frac{p\gamma}{8}\psi_{1,\mathbf{k},2}
\end{aligned}
\end{equation} 
\begin{equation} \label{stabsystemslow4}
\begin{aligned}
&\frac{d\psi_{2,\mathbf{k},2}}{dt}=(-{D}_tk^2-4-2i{\mu}_{2}-\frac{p\gamma}{8})\psi_{2,\mathbf{k},2}\\&
+\frac{p\gamma}{8}\psi_{2,\mathbf{k},-2}+\frac{p\gamma}{8}\psi_{1,\mathbf{k},-2}-\frac{p\gamma}{8}\psi_{1,\mathbf{k},2}
\end{aligned}
\end{equation} 
This system has a characteristic polynomial whose solutions we can find exactly.  For simplicity we can assume $\mu_1$ and $\mu_2$ are equal in magnitude ($|{\mu}|$) and opposite in direction. 
From computations on mathematica we find the eigenvalues of this system. In particular we find that two of the four eigenvalues are always with negative real parts and the third and fourth eigenvalues' real parts are given by $-D_tk^2-4-\frac{p\gamma}{4}\pm 2\sqrt{-\mu^2+\frac{p^2\gamma^2}{64}}$. These two eigenvalue's real parts are plotted in figure \ref{two eigenvalues}. There we see that the angular velocity has a stabilizing effect on the dominant eigenvalue. {This shows that increasing angular velocity has a stabilizing effect on our uniform isotropic state for the no swimming (rotor) parameter regime. Note that the stabilizing and decorrelating effect of translational swimming in an active suspension of pusher microswimmers had been analytically studied in \citep{ohm1, Morozov1}}

\section{Correlations} \label{Correlations}
In this section we will look at the spatial autocorrelations of the suspension. Like in the previous section we will consider different parameter regimes case by case and in each case go over different types of correlations. First we will derive of the matrix equation for the phase space's Fourier mode's covariance matrix and define the correlations that will be considered.

To find the correlation matrix equation we start with the phase space equation in Fourier space:
\begin{equation} \label{Fourier_phase_space_SDE}
\begin{aligned}
        d\hat{\psi}_{\mathbf{k}}=\mathbf{L}_k\psi_kdt+d\tilde{W}_{\mathbf{k},l,t}
\end{aligned}
\end{equation}
with the vectors' components giving the equations for the angular modes. We can formally treat the system as an Ornstein-Uhlenbeck process and write the solution as:
\begin{equation} \label{Fourier_phase_space_SDE2}
\begin{aligned}
        \hat{\psi}_{\mathbf{k}}=\exp(\mathbf{L}_k[t])\hat{\psi}_{0,k}+\int\limits_{0}^{t}\exp(\mathbf{L}_k[t-s])d\tilde{W}_{\mathbf{k},l,s}
\end{aligned}
\end{equation}
We then define the spatial covariance matrix to be $X(t)=\langle \hat{\psi}_{\mathbf{k}}(t)\hat{\psi}_{\mathbf{k}}(t)^H \rangle-\langle\hat{\psi}_{\mathbf{k}}(t)\rangle\langle\hat{\psi}_{\mathbf{k}}(t)\rangle^H $. 
\begin{equation} \label{Fourier_phase_space_Corr}
\begin{aligned}
       &X(t)=\int\limits_0^t\exp(\mathbf{L}_k[t-s])\langle d\tilde{W}_{\mathbf{k},l,s}d\tilde{W}^H_{\mathbf{k},l',s}\rangle\exp(\mathbf{L}^H_k[t-s])\\&
       X(t)=\int\limits_0^t\exp(\mathbf{L}_k[t-s])(\frac{2}{N_p}[k^2D_t+l^2]\delta_{l,l'})\exp(\mathbf{L}^H_k[t-s])ds
\end{aligned}
\end{equation} 
Differentiating with Leibniz integral rule we have:
\begin{equation} \label{Fourier_phase_space_Corr2}
\begin{aligned}
       \frac{dX(t)}{dt}=(\frac{2}{N_p}[k^2D_t+l^2]\delta_{l,l'})+\mathbf{L}_kX(t)+X(t)\mathbf{L}_k^H
\end{aligned}
\end{equation} 
Making the reasonable assumption that $\lim\limits_{t \rightarrow \infty} X(t) $ exists we get following equation for the stationary solution of the Covariance matrix X:
\begin{equation} \label{Fourier_phase_space_Corr3}
\begin{aligned}
       0=(\frac{2}{N_p}[k^2D_t+l^2]\delta_{l,l'})+\mathbf{L}_kX+X\mathbf{L}_k^H
\end{aligned}
\end{equation} 
We can assign $\hat{M}=(\frac{2}{N_p}[k^2D_t+l^2]\delta_{l,l'})$ and we get:
\begin{equation} \label{Fourier_phase_space_Corr4}
\begin{aligned}
       \mathbf{L}_kX+X\mathbf{L}_k^H=-\hat{M}
\end{aligned}
\end{equation} 
The above equation is known as the continuous Lyapunov equation in control theory. 

\subsection{Definitions and Numerical Results}
In this section we will give the definitions of the autocorrelations of the suspension that we will discuss in detail and given their matrix based results. The numerical results are derived from a 150x150 $\mathbf{L}_k$ matrix where the two diagonal 75x75 matrices stand for the two spin sub-populations. The software used is MATLAB.

\subsubsection{Concentration Correlation}
\begin{figure}[h!]	 
\includegraphics[scale=0.55]{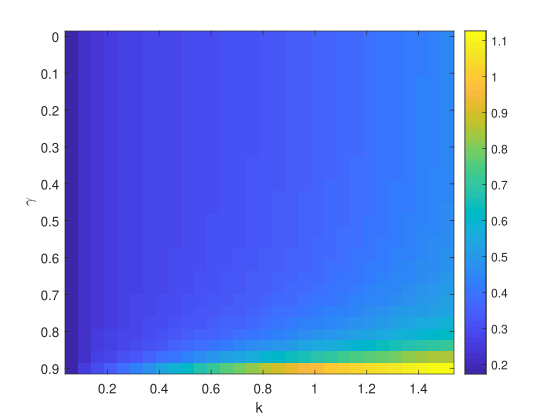}
\par
\includegraphics[scale=0.55]{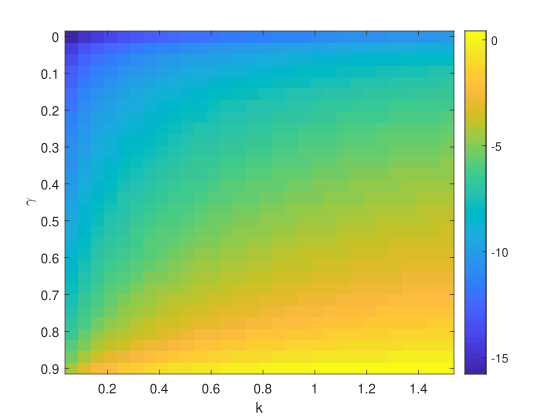}
\caption{ $\log(N_pC_{c,k})$ for same spin (top) and $\log(-N_pC_{c,k})$ opposite spin (bottom) populations with shape parameter($\gamma$) and wavenumber(k). The non-dimensional parameters were as follows: $N_p=10^6$, $|\mu|=1$, $\nu=1$, $p=10$, and $D_t=0.01$}
\label{conc_corr_fig}
\end{figure}

Grouping of swimmers is an important indicator of emergent behavior in suspensions. This grouping or clustering can be measured by the spatial concentration correlation. The concentration is given by:
\begin{equation} \label{conc_defn}
\begin{aligned}
C(\boldsymbol{x},t)=\frac{1}{2\pi}\int \psi(\boldsymbol{x},n,t)\mathbf{dn}
\end{aligned}
\end{equation}
Then, the spatial concentration correlation between two points $\boldsymbol{x}$ and $\mathbf{x'}$ is given by:
\begin{equation} \label{conc_corr_defn}
\begin{aligned}
C_c(\mathbf{r})=\frac{1}{L^2}\int\limits_{0x0}^{LxL}dx\int\limits_{0x0}^{LxL} dx' \langle C(\boldsymbol{x},t)C(\mathbf{x'},t)\rangle \delta(\mathbf{r-x+x'})
\end{aligned}
\end{equation}
and its Fourier coefficient is: 
\begin{equation} \label{conc_corr_Fourier}
\begin{aligned}
C_{c,\textbf{k}}=\langle \psi^*_{\textbf{k},0} \psi_{\textbf{k},0}\rangle
\end{aligned}
\end{equation}
Figure \ref{conc_corr_fig} gives the concentration correlations for same spin and opposite spin populations. It shows us that we get a positive concentration correlation for same spin populations and negative concentration correlation for opposite spin populations under our parameter regime. This is consistent with the clustering phenomena observed with numerical experiments for micro-rotors in denser suspensions \cite{Yeo2015}. 

\subsubsection{Orientation Correlation}
A suspension is nematic if its individual swimmers tend to align their orientation in a given direction. The nematic nature of the suspension can be measured by the spatial orientation correlation. The orientation is given by:
\begin{equation} \label{conc_defn2} 
\begin{aligned}
\mathbf{N}(\boldsymbol{x},t)=\frac{1}{2\pi}\int \mathbf{n}\psi(\boldsymbol{x},n,t)\mathbf{dn}
\end{aligned}
\end{equation}
Then, the spatial orientation correlation between two points $\boldsymbol{x}$ and $\mathbf{x'}$ is given by:
\begin{equation} \label{orn_corr_defn}
\begin{aligned}
C_o(\mathbf{r})=\frac{1}{L^2}\int\limits_{0x0}^{LxL}dx\int\limits_{0x0}^{LxL} dx' \langle \mathbf{N}(\boldsymbol{x},t)\mathbf{N}(\mathbf{x'},t)\rangle \delta(\mathbf{r-x+x'})
\end{aligned}
\end{equation}
and its Fourier coefficient is: 
\begin{equation} \label{orn_corr_Fourier}
\begin{aligned}
C_{o,\textbf{k}}=\frac{1}{2}\langle \psi^*_{\textbf{k},1} \psi_{\textbf{k},1}+\psi^*_{\textbf{k},-1} \psi_{\textbf{k},-1}\rangle
\end{aligned}
\end{equation}

\begin{figure}[h!]	 
\includegraphics[scale=0.55]{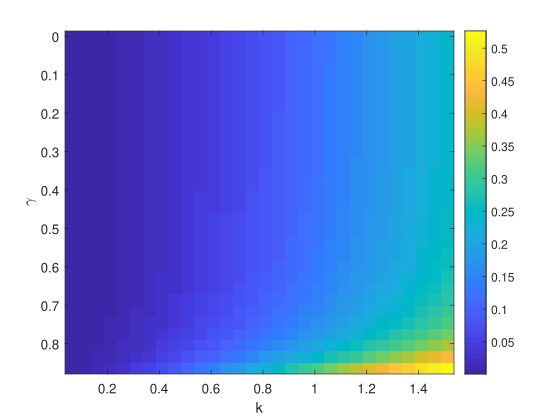}
\par
\includegraphics[scale=0.55]{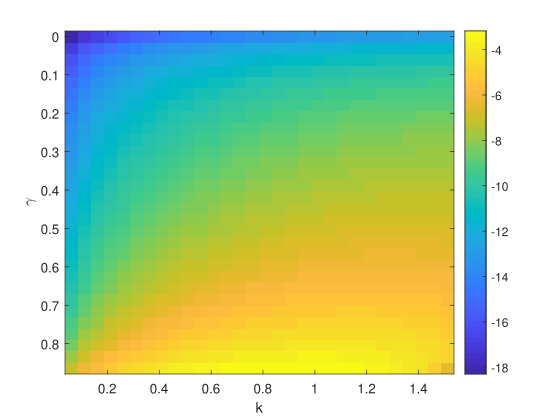}
\caption{ $\log(N_pC_{o,k})$ for same spin (top) and $\log(-N_pC_{o,k})$ opposite spin (bottom) populations with shape parameter($\gamma$) and wavenumber(k). The non-dimensional parameters were as follows: $N_p=10^6$, $|\mu|=1$, $\nu=1$, $p=10$, and $D_t=0.01$}
\label{orn_corr_fig}
\end{figure}

Figure \ref{orn_corr_fig} gives the orientation correlations for same spin and opposite spin populations. It shows us that we get a positive orientation correlation for same spin populations and negative orientation correlation for opposite spin populations under our parameter regime. {The orientation correlation of a micro-rotor pair due to optic effects has been experimentally observed and modeled using Faxen's law in \citep{Benzion}. There they observed positive correlations for a pair of same spin micro-rotors.}

\subsubsection{Fluid active stress Correlation}
\begin{figure}[h!]	
\includegraphics[scale=0.55]{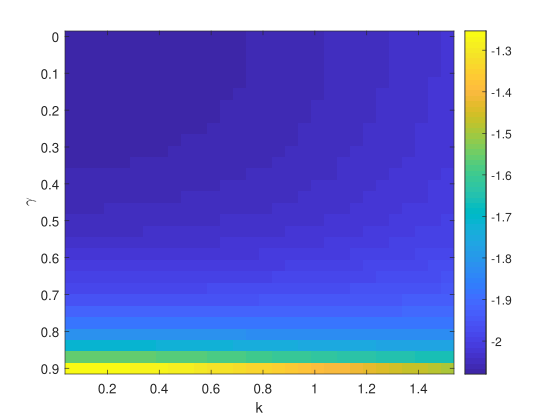}
\par
\includegraphics[scale=0.55]{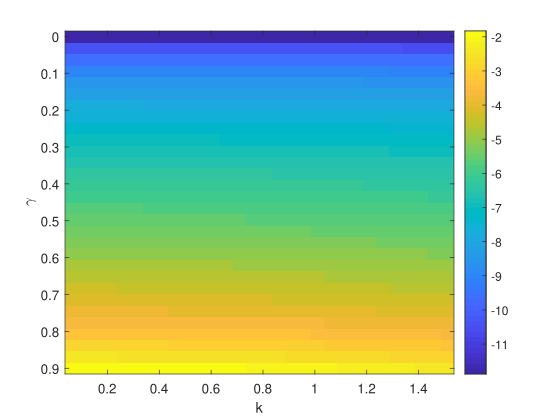}
\caption{ $\log(N_pC_{s,k})$ for same spin (top) and $\log(-N_pC_{s,k})$ opposite spin (bottom) populations with shape parameter($\gamma$) and wavenumber(k). The non-dimensional parameters were as follows: $N_p=10^6$, $|\mu|=1$, $\nu=1$, $p=10$, and $D_t=0.01$}
\label{stress_corr_fig}
\end{figure}

Active stress of the fluid measures the force exerted in the fluid by the active particles. The active stress is given by:
\begin{equation} \label{stress_defn} 
\begin{aligned}
\mathbf{\Sigma}(\boldsymbol{x},t)=&\frac{p}{2\pi}\int \psi(\boldsymbol{x},n,t)(\mathbf{nn}-\frac{1}{2}\mathbf{I})\mathbf{dn}\\\end{aligned}
\end{equation}
The spatial active stress correlation between two points $\boldsymbol{x}$ and $\mathbf{x'}$ is given by:
\begin{equation} \label{stress_corr_defn}
\begin{aligned}
C_s(\mathbf{r})=\frac{1}{L^2}\int\limits_{0x0}^{LxL}dx\int\limits_{0x0}^{LxL} dx' \langle \mathbf{\Sigma}(\boldsymbol{x},t)\mathbf{\Sigma}(\mathbf{x'},t)\rangle \delta(\mathbf{r-x+x'})
\end{aligned}
\end{equation}
The shear stress autocorrelation is defined as:
\begin{equation} \label{conc_corr_Fourier2}
\begin{aligned}
C_{s,\textbf{k}}=&\frac{1}{16}\langle -\psi_{\textbf{k},2} \psi^*_{\textbf{k},-2}-\psi_{\textbf{k},-2} \psi^*_{\textbf{k},2}\\&
+\psi_{\textbf{k},2} \psi^*_{\textbf{k},2}+\psi_{\textbf{k},-2} \psi^*_{\textbf{k},-2}\rangle
\end{aligned}
\end{equation}
Figure \ref{stress_corr_fig} gives the shear stress correlations for same spin and opposite spin populations. Here too we see that a same spin populations are positively correlated and opposite spin populations are negatively correlated under our parameter regime. 

Shear stress correlations tell us the correlations of frictional force on the fluid as we spatial vary the swimmer suspension. They can also be used to get Fluid velocity Correlations through the Stokes' equation and from there the effective transport and diffusion properties of tracers. 

\subsection{Perturbative Analysis}

Next we look at perturbations in the linearly stable regime.
\subsubsection{Active Particles with Small $p\gamma$ and Small speed ($\nu$):} 

We assume we have $p\gamma \ll 1$. We get the following $L=L_1+\epsilon  L_2$ and $M=M^{(S)}_1+\epsilon  M^{(S)}_2$ where \\
\begin{equation}
\begin{aligned}
    & L_j=\begin{pmatrix}  L_{j,1} && 0 \\ 0 &&  L_{j,2} \end{pmatrix} \text{   and   }  
    M^{(S)}_j= \begin{pmatrix}  0 && M_{j} \\ M_{j} &&  0 \end{pmatrix}
\end{aligned}
\end{equation}
The block matrices are given by:
\begin{equation}
\begin{aligned}
    & L_{1,\bar{i}}= (-l^2-\mathbf{k}^2D_t-il\mu_{\bar{i}})\delta_{l,l'} 
    \end{aligned}
\end{equation}
\begin{equation}
\begin{aligned}
     L_{2,\bar{i}}=& (-ik e^{i\theta_k})\delta_{l,l'-1}+(-ik e^{-i\theta_k})\delta_{l,l'+1}\\&
     +\delta_{l,2}\tilde{\xi}_1(-\delta_{l',2}+e^{-i4\theta_{k}}\delta_{l',-2})\\&
+\delta_{l,-2}\tilde{\xi}_1(\delta_{l',2}e^{i4\theta_{k}}-\delta_{l',-2})
    \end{aligned}
\end{equation}
\begin{equation}
\begin{aligned}
& M_{1}=\mathbf{0}
\end{aligned}
\end{equation}
\begin{equation}
\begin{aligned}
M_{2}=& \delta_{l,2}\tilde{\xi}_1(-\delta_{l',2}+e^{-i4\theta_{k}}\delta_{l',-2})\\&
+ \delta_{l,-2}\tilde{\xi}_1(\delta_{l',2}e^{i4\theta_{k}}-\delta_{l',-2}) 
\end{aligned}
\end{equation}
Here $\epsilon =\nu$ and $\tilde{\xi}_1=\frac{p\gamma}{8\nu}$ is the non dimensional parameter for stresslet strength/shape parameter. 

The Lyapunov equation at leading order for the diagonal block matrices of the correlation matrix is given by:
\begin{equation}
\begin{aligned}
  L_{1,\bar{i}}X_{1,\bar{i}}+X_{1,\bar{i}}L_{1,\bar{i}}^H=-\hat{M}
\end{aligned}
\end{equation}
Where 
\begin{equation}
\begin{aligned}
  \hat{M}=\frac{2}{N_p}k^2D_t\delta_{l,l'}+\frac{2}{N_p}l^2\delta_{l,l'}
\end{aligned}
\end{equation}
Next we derive the leading order correlation matrix. First we diagonalize $ L_{1,\bar{i}}$. In this leading order case that simply means we leave $ L_{1,\bar{i}}$ as is since its already a diagonal matrix. We then get the following expressions for each diagonal component of either side of the Lyapunov equation (Starting here we will refer to the $(i,j)^{th}$ entry matrix $X_{1,\bar{i}}$ as $X_{1,\bar{i}}[i,j]$ and will use the same notation for other matrices.):
\begin{equation}
\begin{aligned}
  (-2l^2-2\mathbf{k}^2D_t)X_{1,\bar{i}}[j,j]=-\hat{M}[j,j]
\end{aligned}
\end{equation}
The off-diagonal components will be zero due to the RHS. $X_{1,\bar{i}}$ is now shown to be $\frac{1}{N_p}I$. We notice that the correlations (including concentration and orientation correlation) will be invariant with the advent of active angular velocity at leading order. Next we will find the first order correlation matrix.
\begin{equation}
\begin{aligned}
  L_{1,\bar{i}}X_{2,\bar{i}}+X_{2,\bar{i}}L_{1,\bar{i}}^H=-(L_{2,\bar{i}})X_{1,\bar{i}}-X_{1,\bar{i}}(L_{2,\bar{i}})^H
\end{aligned}
\end{equation}

From here we get the equations for the entries of $X_2$.
\begin{equation}
\begin{aligned}
  (X_{2,\bar{i}})[l,l']&=
  -\frac{1}{N_p}[\delta_{l,2}\tilde{\xi}_1(-2\delta_{l',2}+2e^{-i4\theta_{k}}\delta_{l',-2})\\&
+\delta_{l,-2}\tilde{\xi}_1(2e^{i4\theta_{k}}\delta_{l',2}-2\delta_{l',-2})]\\&
\cdot[(-l^2-k^2D_t-i\mu_{\bar{i}} l)\\&
+(-(l')^2-k^2D_t+i\mu_{\bar{i}} l')]^{-1}
\end{aligned}
\end{equation}
Studying the correction to the correlation matrix shows that there is no correction for the concentration correlation and orientation correlations but there is a correction for stress correlation involving the angular velocities . In particular, we get the following stress auto-correlation:
\begin{equation}
\begin{aligned}
  C^{(\text{in})}_{s,\mathbf{k}}&=\frac{1}{8N_p}
  -\frac{p\gamma}{64N_p}\Big(\frac{1}{4+k^2D_t}\\&
  -\frac{\cos(4\theta_k)(4+k^2D_t)-2\sin(4\theta_k)\mu_{\bar{i}}}{(4+k^2D_t)^2+4\mu^2_{\bar{i}}}  \Big)
\end{aligned}
\end{equation}
For simplicity we set $\theta_k=0$ and get:
\begin{equation}
\begin{aligned}
  C^{(\text{in})}_{s,\mathbf{k}}|_{\theta_k=0}&=\frac{1}{8N_p}
  -\frac{p\gamma}{64N_p}\Big(\frac{1}{4+k^2D_t}\\&
  -\frac{(4+k^2D_t)}{(4+k^2D_t)^2+4\mu^2_{\bar{i}}}  \Big)
\end{aligned}
\end{equation}
Here we notice that the angular velocity has a positive impact in creating positive stress correlations for within population swimmer-rotors. Note that $X_{2,\bar{i}}$ consists of the diagonal blocks of the correction correlation matrix. Next we work out the off-diagonal blocks of the $X_2$ matrix to get corrections relevant for the cross-population auto-correlations. We will call the top right non-diagonal block matrix $X_{2,TR}$ and the bottom left non-diagonal block matrix $X_{2,BL}$. This notation for non-diagonal block matrices will be used for other matrices as well in this paper. 
\begin{equation}
\begin{aligned}
  L_{1,1}X_{2,TR}+X_{2,TR}L_{1,2}^H=-(M_{2})(\frac{I}{N_p})-(\frac{I}{N_p})(M_{2})^H
\end{aligned}
\end{equation}
\begin{equation}
\begin{aligned}
  L_{1,2}X_{2,BL}+X_{2,BL}L_{1,1}^H=-(M_{2})(\frac{I}{N_p})-(\frac{I}{N_p})(M_{2})^H
\end{aligned}
\end{equation}
\begin{equation}
\begin{aligned}
  &(X_{2,TR})[l,l']= -\frac{1}{N_p}\Big(\delta_{l,2}\tilde{\xi}_1(-2\delta_{l',2}+2e^{-i4\theta_{k}}\delta_{l',-2})\\&
+\delta_{l,-2}\tilde{\xi}_1(2e^{i4\theta_{k}}\delta_{l',2}-2\delta_{l',-2})  \Big)\\& [(-l^2-k^2D_t-i\mu_{1} l)+(-(l')^2-k^2D_t+i\mu_{2} l')]^{-1}
\end{aligned}
\end{equation}
\begin{equation}
\begin{aligned}
  &(X_{2,BL})[l,l']= -\frac{1}{N_p}\Big(\delta_{l,2}\tilde{\xi}_1(-2\delta_{l',2}+2e^{-i4\theta_{k}}\delta_{l',-2})\\&
+\delta_{l,-2}\tilde{\xi}_1(2e^{i4\theta_{k}}\delta_{l',2}-2\delta_{l',-2})  \Big)\\& [(-l^2-k^2D_t+i\mu_{1} l)+(-(l')^2-k^2D_t-i\mu_{2} l')]^{-1}
\end{aligned}
\end{equation}
From {these} expressions we see that there are no corrections for the cross population concentration and orientation correlations but that the cross population stress correlations are corrected by a term involving angular velocities (Here the angular velocity has a negative impact in creating negative stress correlations when $\theta_k=0$). The cross population stress auto-correlation is given by: 
\begin{equation}
\begin{aligned}
  &\frac{1}{16}\langle -X_{\mathbf{k},2}(t)X^*_{\mathbf{k},n-2}(t,t+\Delta t)-X_{\mathbf{k},-2}(t)X^*_{\mathbf{k},n+2}(t,t+\Delta t)\\&
  +X_{\mathbf{k},2}(t)X^*_{\mathbf{k},n+2}(t,t+\Delta t)+X_{\mathbf{k},-2}(t)X^*_{\mathbf{k},n-2}(t,t+\Delta t) \rangle\\&
  \approx -\frac{p\gamma}{64N_p}\Big(\frac{8+2k^2D_t}{(4+k^2D_t)^2+(\mu_2-\mu_1)^2}\\&
  -\frac{\cos(4\theta_k)(4+k^2D_t)+ ie^{i4\theta_k}(\mu_{1}-\mu_{2})}{(4+k^2D_t)^2+(\mu_{1}+\mu_{2})^2}  \Big) 
\end{aligned}
\end{equation}

We see that same spin stress correlation is positive and decreasing in wave-number magnitude $k$ while the different spin stress correlation is negative and decreasing in k for $\theta_k=0$. We also see a linear relationship with the shape parameter $\gamma$ in both of these cases. These observations agree with the general numerical results on figure \ref{stress_corr_fig}.

\subsubsection{Active Particles with Small $p\gamma$ and Large Angular Speed ($|\mu_m|$):}
The {second} method we will use to find the nearly circular small $p\gamma$  correlations is to take the original particle level SDEs at (\ref{eq_c}) and (\ref{eq_d}) and homogenize them with a long time assumption. The assumption of long time amounts to a stability requirement and the assumption of small rotational diffusion. We rewrite the equations here:
\begin{equation} 
    \begin{aligned}
    d\mathbf{\tilde{x}_m}(t)=&(\mathbf{u_m}(t)+\nu(\cos(\theta_m),\sin(\theta_m)))dt\\&
    +\sqrt{2D_t}d\mathbf{W_{x,m,t}}
    \end{aligned}
\end{equation}
\begin{equation} 
    \begin{aligned}
    d\theta_m(t)=(\omega_m(t)+{\mu}_m )dt+\sqrt{2}d\mathbf{W_{\theta,m,t}}
    \end{aligned}
\end{equation}
In the case of small $\frac{p\gamma}{8}$ the flow (u) and vorticity ($\omega$) terms are small. With $\epsilon =\frac{p\gamma}{8}$ and the $\gamma$ factor getting factored out of u and $\omega$, we write $u=\epsilon \tilde{u}$ and $\omega=\epsilon \tilde{\omega}$.
 We then let $\breve{t}=\hat{\epsilon }^2 t$ and $\breve{x}_m=\hat{\epsilon }\tilde{x}_m$. We then get:
\begin{equation} 
    \begin{aligned}
    d\breve{\boldsymbol{x}}_m(\breve{t})=&\frac{1}{\hat{\epsilon }}(\epsilon \big[{\mathbf{\tilde{u}_m}(\breve{t})}\big]+\nu(\cos(\theta_m),\sin(\theta_m)))d\breve{t}\\&
    +{\sqrt{2D_t}}d\mathbf{W_{x_m,t}}
    \end{aligned}
\end{equation}
\begin{equation} 
    \begin{aligned}
    d\theta_m(\breve{t})=\frac{1}{\hat{\epsilon }^2}(\epsilon \omega_m(\breve{t})+\hat{\mu}_m )d\breve{t}+\frac{\sqrt{2}}{\hat{\epsilon }}d\mathbf{W_{\theta_m,t}}
    \end{aligned}
\end{equation}
Assuming that $\epsilon $ is small we drop its corresponding terms to get: 
\begin{equation} 
    \begin{aligned}
    d\breve{\boldsymbol{x}}_m(\breve{t})=\frac{\nu}{\hat{\epsilon }}((\cos(\theta_m),\sin(\theta_m)))d\breve{t}+{\sqrt{2D_t}}d\mathbf{W_{x_m,t}}
    \end{aligned}
\end{equation}
\begin{equation} 
    \begin{aligned}
    d\theta_m(\breve{t})=\frac{1}{\hat{\epsilon }^2}(\hat{\mu}_m )d\breve{t}+\frac{\sqrt{2}}{\hat{\epsilon }}d\mathbf{W_{\theta_m,t}}
    \end{aligned}
\end{equation}
We have a standard slow-fast scale separation with the micro-swimmer being a fast rotor and a slow swimmer. N.b. this is not due to $\nu$, the translational velocity, being small but due to our formal small parameter $\hat{\epsilon }$. Averaging out $\theta_m$ from $\mathbf{\breve{x}}_m$ we notice that the centering condition is met. Hence, homogenization is used to simplify the system. The homogenization will give the position as:
\begin{equation} 
    \begin{aligned}
    d\breve{\boldsymbol{x}}_m(\epsilon ^{-2}\breve{t})=\big({2D_t}+\big(\frac{ \nu}{1+\mu_m^2}\big)\big)^{\frac{1}{2}}d\mathbf{W_{x_m,t}}
    \end{aligned}
\end{equation}
Rescaling we get:
\begin{equation} 
    \begin{aligned}
    &d\boldsymbol{x}_m(t)=\big({2D_t}+\big(\frac{\nu}{1+\mu_m^2}\big)\big)^{\frac{1}{2}}d\mathbf{W_{x_m,t}}\\&
     d\theta_m(t)=(\hat{\mu}_m )dt+\sqrt{2}d\mathbf{W_{\theta_m,t}}
    \end{aligned}
\end{equation}
Let $\breve{D}_t=D_t+\frac{1}{2}\big(\frac{\nu}{1+\mu_m^2}\big)$. For our application $\mu_m$ has two types (opposite spin).\\
Now we go through the same process process as before with the construction of the SPDE for the suspension and its linearization. We split the angular velocity terms into two for the two sub populations. After Fourier transformation, we get the leading order matrix $L_{1,\bar{i}}$.   
\begin{equation} 
    \begin{aligned}
(L_{1,\bar{i}})_{l,l'}=(-\breve{D}_tk^2-l^2-\hat{i}\mu_{\bar{i}}l)\delta_{l,l'}
    \end{aligned}
\end{equation}
 The Lyapunov Equation at leading order for the diagonal block matrices of the correlation matrix is given by:
\begin{equation} 
    \begin{aligned}
 L_{1,\bar{i}}X_{1,\bar{i}}+X_{1,\bar{i}}(L_{1,\bar{i}})^H=-\hat{M}_1
     \end{aligned}
\end{equation}
Where $\hat{M}_1=(\frac{2}{N_p}\breve{D}_tk^2+\frac{2}{N_p}l^2)\delta_{l,l'}$. Since $L_1$ is a diagonal matrix we get:
 \begin{equation} 
    \begin{aligned}
(X_{1,\bar{i}})_{l,l'}=\frac{-(\hat{M}_1)_{l,l'}}{L_{1,\bar{i}}^{l,l}+\bar{L}_{1,\bar{i}}^{l',l'}}
     \end{aligned}
\end{equation}
Since $(L_{1,\bar{i}})_{l,l}+(\bar{L}_{1,\bar{i}})_{l',l'}=-2(\breve{D}_tk^2+l^2)$, $X_{1,\bar{i}}=\frac{1}{N_p}I$. The off-diagonal block matrices of $X_1$ are zero matrices. We no significant correlations here so we move to the next order. $L_2$ is found by subtracting $L_1$ from the matrix we would get without the initial homogenization. First we have:
 \begin{equation} 
    \begin{aligned}
  \tilde{L}_{2,\bar{i}}=&-(D_t-\breve{D}_t)k^2\delta_{l,l'}+(-i\nu ke^{i\theta_k})\delta_{l,l'-1}\\&
  +(-i\nu ke^{-i\theta_k})\delta_{l,l'+1}+\delta_{l,2}(\delta_{l',-2}\epsilon  e^{-4\theta_k}-\delta_{l',2}\epsilon )\\&
  +\delta_{l,-2}(\delta_{l',2}\epsilon e^{4\theta_k}-\delta_{l',-2}\epsilon ) 
     \end{aligned}
\end{equation}
Note that $\tilde{L}_{2,\bar{i}}$ is a smaller matrix compared to $L_{1,\bar{i}}$ (in terms of the reduced 5x5 matrix) when the ratio below is small.
 \begin{equation} 
    \begin{aligned}
    \frac{||\tilde{L}_2||_1}{||L_1||_1}=\frac{|D_t-(\breve{D}_t)_{max}|k^2+2\nu k+4\epsilon}{([(\breve{D}_t)_{max}k^2+4]^2+4\mu_{max}^2)^\frac{1}{2}}=\tilde{\epsilon}
        \end{aligned}
\end{equation}

Among other things $\tilde{\epsilon}$ is small when $\mu>>1$. Hence this case can be consider the case with high active angular velocity. We rewrite the L matrix as follows:
\begin{equation}
    \begin{aligned}
L_{\tilde{i}}=L_{1,\tilde{i}}+\tilde{\epsilon}L_{2,\tilde{i}}  \text{   where } L_{2,\tilde{i}}=\frac{1}{\tilde{\epsilon}}\tilde{L}_{2,\tilde{i}}
    \end{aligned}
\end{equation}
When $\tilde{\epsilon}$ is small the approach is expected to work. Using this we get the following higher order equation for the diagonal block matrices of the correlation matrix: 
\begin{equation} 
    \begin{aligned}
 L_{1,\bar{i}}X_{2,\bar{i}}+X_{2,\bar{i}}(L_{1,\bar{i}})^H=-L_{2,\bar{i}}X_{1,\bar{i}}-X_{1,\bar{i}}L_{2,\bar{i}}^H
     \end{aligned}
\end{equation}
\begin{equation} 
    \begin{aligned}
 &L_{1,\bar{i}}X_{2,\bar{i}}+X_{2,\bar{i}}(L_{1,\bar{i}})^H=
 -\frac{2}{\tilde{\epsilon}N_p}\big[-(D_t-\breve{D}_t)k^2\delta_{l,l'}\\&+\delta_{l,2}(\delta_{l',-2}\epsilon e^{-4\theta_k}-\delta_{l',2}\epsilon )
 +\delta_{l,-2}(\delta_{l',2}\epsilon e^{4\theta_k}-\delta_{l',-2}\epsilon ) \big]
     \end{aligned}
\end{equation}
\begin{equation} 
    \begin{aligned}
&(X_{2,\bar{i}})_{l,l'}=-\big(\frac{2}{N_p}\big)\frac{1}{L_{1,\bar{i}}^{l,l}+\bar{L}_{1,\bar{i}}^{l',l'}}\big[-(D_t-\breve{D}_t)k^2\delta_{l,l'}\\&
+\delta_{l,2}(\delta_{l',-2}\epsilon e^{-4\theta_k}-\delta_{l',2}\epsilon )+\delta_{l,-2}(\delta_{l',2}\epsilon e^{4\theta_k}-\delta_{l',-2}\epsilon ) \big]\\&
\cdot\big(\frac{([\breve{D}_tk^2+4]^2+4\mu_{max}^2)^\frac{1}{2}}{|D_t-\breve{D}_t|k^2+2\nu k+4\epsilon}\big)
     \end{aligned}
\end{equation}
\begin{equation} 
    \begin{aligned}
&(X_{2,\bar{i}})_{l,l'}=\big(\frac{1}{N_p}\big)\frac{1}{(\breve{D}_tk^2+l^2)}\big[-(D_t-\breve{D}_t)k^2\delta_{l,l'}\\&
+\delta_{l,2}(\delta_{l',-2}\epsilon e^{-4\theta_k}-\delta_{l',2}\epsilon )+\delta_{l,-2}(\delta_{l',2}\epsilon e^{4\theta_k}-\delta_{l',-2}\epsilon ) \big]\\& \cdot\big(\frac{([\breve{D}_tk^2+4]^2+4\mu_{max}^2)^\frac{1}{2}}{|D_t-\breve{D}_t|k^2+2\nu k+4\epsilon}\big)
     \end{aligned}
\end{equation}
From this equation and the previous leading order result we gather the enhanced same spin concentration correlation:
$C_{c,\mathbf{k}}=\frac{1}{N_p}+\frac{1}{N_p}(\frac{\breve{D}_t-D_t}{\breve{D}_t})$, the enhanced same spin orientation correlation: $C_{o,\mathbf{k}}=\frac{1}{N_p}+\frac{1}{N_p}(\frac{(\breve{D}_t-D_t)k^2}{\breve{D}_tk^2+1})$, and the enhanced same spin stress correlation: $C_{s,\mathbf{k}}=\frac{1}{N_p}+\frac{1}{16N_p}(\frac{2(\breve{D}_t-D_t)k^2-\epsilon(2+e^{-4\theta_k}+e^{4\theta_k})}{\breve{D}_tk^2+4})$. 

We notice that both the same spin concentration correlation and orientation correlation are positive and grow with speed $\nu$ and decline with angular speed $\mu$. For same spin stress correlation we again see that the speed has a monotonically positive relationship and that angular speed has a monotonically negative relationship.

\section{Discussion}  \label{discussion}
The collective mobility properties of active particles have been a subject of extensive investigation. We continued this investigation by looking at the uniform suspension stability and spatial correlations of hydrodynamically interacting microswimmers and microrotors. We made sure to have a heterogeneous (bi-modal) collection in terms of angular velocity.

Generally speaking we see that angular speed has a stabilizing and de-correlating effect. We saw that the angular speed has a stabilizing effect on suspensions with naturally occurring parameter regimes of interest. We also saw that, for particles with the same angular velocity an increase in angular speed leads to a decrease in concentration, orientation, and stress spatial correlations when we have small particle translational speed and small suspension number density relative to the product of the viscosity, inverse rotational diffusion time scale and inverse stresslet coefficient ($\frac{\eta \tilde{D}_r L^2}{N_p}$) or when we have large particle angular speed and small suspension number density relative to $\frac{\eta \tilde{D}_r L^2}{N_p}$. We also showed evidence through our matrix based computations that the same spin microrotors have positive concentration and orientation correlation while opposite spin microrotors have negative concentration and orientation correlation. This helps present evidence for segregated pattern formation in multi-spin suspensions of microrotors. {It is important to note here that pattern formation in active suspensions beyond the linearly stable regime (on which this paper is based) is a recent subject of study \citep{ohm2}.}

These insights on the collective statistical and dynamic properties of the microswimmers and microrotors can be readily extended for other parameter regimes of interest like is done in the appendix. 

\section{Acknowledgements}
The author thanks Dr. Peter R. Kramer and Dr. Patrick T. Underhill. for the advice given concerning stochastic kinetic theory and suspension microhydrodynamics.

\section{Appendix A: Stability Analysis for Other Parameter Regimes}

Another approach for small $p\gamma$ is to consider the angular velocity to be order one while all other terms are small. This case is where we clearly saw instability for small $p\gamma$ from the matrix based numerical computations. In particular we see concentration correlation leading to an instability. In this case we can perform a regular perturbation to get the asymptotic approximation of the spectrum.
\begin{equation}
L=\bar{L}+\epsilon  M \text{ where } \bar{L}_{\textbf{k},l,l'}=\begin{pmatrix} a_{\textbf{k},l,l'} && 0 \\
0 && d_{\textbf{k},l,l'}\end{pmatrix},
\end{equation}
\begin{equation}
a_{\textbf{k},l,l'}=-{il\mu_1}\delta_{l,l'} \text{   }b_{\textbf{k},l,l'}=-{il\mu_2}\delta_{l,l'}
\end{equation}
\begin{equation}
\text{and } M=\begin{pmatrix} \alpha_{\textbf{k},l,l'} && \beta_{\textbf{k},l,l'}\\
\gamma_{\textbf{k},l,l'} && \zeta_{\textbf{k},l,l'}\end{pmatrix} \text{ with }
\end{equation}
\begin{equation}
\begin{aligned}
\alpha_{\textbf{k},l,l'}=&\zeta_{\textbf{k},l,l'}=\\&
[-\epsilon -\xi_2]\delta_{l,l'}-\xi_3\frac{i}{2}e^{i\theta_{k}}\delta_{l,l'-1}-\xi_3\frac{i}{2}e^{-i\theta_{k}}\delta_{l,l'+1}\\&
+\delta_{l,2}(-\delta_{l',2}+e^{-i4\theta_{k}}\delta_{l',-2})\\&
+\delta_{l,-2}(\delta_{l',2}e^{i4\theta_{k}}-\delta_{l',-2})
\end{aligned}
\end{equation}
\begin{equation}
\begin{aligned}
\beta_{\textbf{k},l,l'}&=\gamma_{\textbf{k},l,l'}=\delta_{l,2}(-\delta_{l',2}+e^{-i4\theta_{k}}\delta_{l',-2})\\&
+\delta_{l,-2}(\delta_{l',2}e^{i4\theta_{k}}-\delta_{l',-2})
\end{aligned}
\end{equation}
Here $\epsilon =\frac{p\gamma}{8}$. $\xi_1 $, $\xi_2$, and $\xi_3$ are $\mathcal{O}(1)$ are non dimensional parameters giving the relationship between the shape parameter and the rotational diffusion, translational diffusion, and the active translational velocity respectively. Namely,\\
\begin{equation}
\begin{aligned}
\xi_1 (k)=\frac{{8D_t}k^2}{p\gamma}, \hspace{0.6 cm}  \xi_2(l)=\frac{{8}l^2}{p\gamma}, \hspace{0.6 cm} \xi_3(k)=\frac{8\nu k}{p\gamma}
\end{aligned}
\end{equation}
The asymptotic approximations of the eigenvalues and eigenvectors are:
\begin{equation}
\begin{aligned}
\lambda_{n}=\bar{\lambda_{n}}+\epsilon \hat{\lambda_{n}} \text{,   }\mathbf{{V}_{n}}=\mathbf{\bar{V}_{n}}+\epsilon \mathbf{\hat{V}_{n}}
\end{aligned}
\end{equation}
\begin{equation}
\begin{aligned}
\bar{\lambda_{n}}=-il{\mu_1} \text{ for } n \leq \frac{N}{2} \text{ and}
\end{aligned}
\end{equation}
\begin{equation}
\begin{aligned}
\bar{\lambda_{n}}=-il{\mu_2} \text{ for } n > \frac{N}{2}
\end{aligned}
\end{equation}
\begin{equation}
\begin{aligned}
\mathbf{\bar{V}_{n}}=(0,0,...,0,1_{n},0,...)
\end{aligned}
\end{equation}
The correction terms are given by:
\begin{equation}
\begin{aligned}
\mathbf{\hat{V}_{n}}=(\bar{\lambda_{n}}I-\bar{L})^{-1}(I-\frac{\mathbf{\bar{V}_{n}\bar{W}_{n}}^*}{\mathbf{\bar{V}_{n}^*\bar{W}_{n}}})M\mathbf{\bar{V}_{n}}
\end{aligned}
\end{equation}
Where ${W}_{n}=\bar{W}_{n}+\epsilon \hat{W}_{n}$ are eigenvectors of $L^*$. It follows that $\bar{W}_{n}$ are eigenvectors of $\bar{L}^*$ since the conjugation will preserve the asymptotic ordering. From this it follows that $\bar{W}_{n}=\bar{V}_{n}$. Hence we get:

\begin{equation}
\begin{aligned}
\hat{\lambda_{n}}=\frac{\mathbf{\bar{V}_{n}}^{*}M\mathbf{\bar{V}_{n}}}{\mathbf{\bar{V}_{n}^*\bar{V}_{n}}}=\mathbf{\bar{V}_{n}}^{*}M\mathbf{\bar{V}_{n}}
\end{aligned}
\end{equation}
\begin{equation}
\begin{aligned}
\mathbf{\hat{V}_{n}}=(\bar{\lambda_{n}}I-\bar{L})^{-1}(I-\mathbf{\bar{V}_{n}\bar{V}_{n}^*})M\mathbf{\bar{V}_{n}}
\end{aligned}
\end{equation}
We notice that $\lambda_{\frac{N}{2}\pm 2}=-il\hat{\mu}_1+\epsilon (-\xi_1 (k)-\xi_2(2)-\xi_3 (k))$. We see that the rotational and translational diffusion as well as the translational speed terms stabilize the system by making the dominant angular modes have negative real eigenvalue parts.

\section{Appendix B: Spatial Correlations for other parameter regimes}

\subsection{Order one $p\gamma$}: 
We get the following $L=L_1+\epsilon  L_2$ and $M=M^{(s)}_1+\epsilon  M^{(s)}_2$ where \\
\begin{equation}
\begin{aligned}
    & L_j=\begin{pmatrix}  L_{j,1} && 0 \\ 0 &&  L_{j,2} \end{pmatrix} \hspace{0.6 cm} \text{and  }  
    M^{(s)}_j= \begin{pmatrix}  0 && M_{j} \\ M_{j} &&  0 \end{pmatrix}
\end{aligned}
\end{equation}
The block matrices are given by:
\begin{equation}
\begin{aligned}
     L_{1,\bar{i}}=& (-l^2-\mathbf{k}^2D_t-il\mu_{\bar{i}})\delta_{l,l'}+\delta_{l,2}\tilde{\xi}_1(-\delta_{l',2}\\&
    +e^{-i4\theta_{k}}\delta_{l',-2})
+\delta_{l,-2}\tilde{\xi}_1(\delta_{l',2}e^{i4\theta_{k}}-\delta_{l',-2}) 
\end{aligned}
\end{equation}
\begin{equation}
\begin{aligned}
& L_{2,\bar{i}}= (-ike^{i\theta_k})\delta_{l,l'-1}+(-ike^{-i\theta_k})\delta_{l,l'+1}
\end{aligned}
\end{equation}
\begin{equation}
\begin{aligned}
& M_{1}=\delta_{l,2}\tilde{\xi}_1(-\delta_{l',2}+e^{-i4\theta_{k}}\delta_{l',-2})\\&
+ \delta_{l,-2}\tilde{\xi}_1(\delta_{l',2}e^{i4\theta_{k}}-\delta_{l',-2})
\end{aligned}
\end{equation}
\begin{equation}
\begin{aligned}
& M_{2}=\mathbf{0}
\end{aligned}
\end{equation}
Here $\epsilon =\nu$ and $\tilde{\xi}_1=\frac{p\gamma}{8}$. The first diagonal block (top left) of the matrix equation is given below. The second diagonal block (bottom right) can be used for our derivations similarly. 
\begin{equation}  \label{slowing swimming corr matrix eqn}
\begin{aligned}
  L_{1,\bar{i}}X_{1,\bar{i}}+M_{1}X_{1,BL}+X_{1,\bar{i}}L_{1,\bar{i}}^H+X_{1,TR}M_{1}^H=-\hat{M}
\end{aligned}
\end{equation}
Where 
\begin{equation}
\begin{aligned}
  \hat{M}=\frac{2}{N_p}k^2D_t\delta_{l,l'}+\frac{2}{N_p}l^2\delta_{l,l'}
\end{aligned}
\end{equation}
Next we derive the leading order correlation matrix. We get the expressions for each diagonal component of either side of the Lyapunov equation. To do this, we first split $L_{1,\bar{i}}$ into its diagonal(minus diagonal stress terms) and non diagonal parts(plus diagonal stress terms). \begin{equation}
\begin{aligned}
  (L_{1,\bar{i}}^{(D)}&+L_{1,\bar{i}}^{(ND)})X_{1,\bar{i}}+X_{1,\bar{i}}(L_{1,\bar{i}}^{(D)}+L_{1,\bar{i}}^{(ND)})^H\\&
  +M_{1}X_{1,BL}+X_{1,TR}M_{1}^H=-\hat{M}
\end{aligned}
\end{equation}
\begin{equation}
\begin{aligned}
  &L_{1,\bar{i}}^{(D)}X_{1,\bar{i}}+L_{1,\bar{i}}^{(ND)}X_{1,\bar{i}}+X_{1,\bar{i}}(L_{1,\bar{i}}^{(D)})^H\\&
  +X_{1,\bar{i}}(L_{1,\bar{i}}^{(ND)})^H+M_{1}X_{1,BL}+X_{1,TR}M_{1}^H=-\hat{M}
\end{aligned}
\end{equation}
$X_{1,BL}=X_{1,TR}^H$ because $X_1$ is a correlation matrix. With $L_{1,\bar{i}}^{(ND)}$ and $M_{1}$ being Hermitian we get
\begin{equation}
\begin{aligned}
  &L_{1,\bar{i}}^{(D)}X_{1,\bar{i}}+X_{1,\bar{i}}(L_{1,\bar{i}}^{(D)})^H+L_{1,\bar{i}}^{(ND)}X_{1,\bar{i}}\\&
  +X_{1,\bar{i}}(L_{1,\bar{i}}^{(ND)})+M_{1}X_{1,BL}+X_{1,BL}^H{^H}M_{1}=-\hat{M}
\end{aligned}
\end{equation}

We find the correlations by diagonalization on the RHS of (\ref{slowing swimming corr matrix eqn}). Doing this will require knowledge of the eigenvectors and eigenvalues of $L_{1}$. We will discuss how to find these conveniently next. 
 First we find the eigensystem of $L_{1,\bar{i}}$. The eigenvalues of the system that excludes the rows that contain non-diagonal entries are simply the diagonal entries and its eigenvector matrix is simply an identity matrix. The rest of $L_{1,\bar{i}}$ forms a 2x2 matrix whose eigenvalues are\\ $\lambda^{(1)}_{1,1}=-\tilde{\xi}_1-4-k^2D_t-\sqrt{\tilde{\xi}_1^2-4\mu_1^2}$ and $\lambda^{(2)}_{1,1}=-\tilde{\xi}_1-4-k^2D_t+\sqrt{\tilde{\xi}_1^2-4\mu_1^2}$.\\
 The eigenvectors are:
 \begin{equation}
\begin{aligned}
    & v_{1}=\begin{pmatrix}  \frac{1}{\tilde{\xi}_1}\big(ie^{4i\Theta_k}(2\mu_1)-e^{4i\Theta_k}\sqrt{\tilde{\xi}_1^2-4\mu_1^2}\big) \\ 1 \end{pmatrix} \text{ and  }  \\&
    v_2= \begin{pmatrix} \frac{1}{\tilde{\xi}_1}\big(ie^{4i\Theta_k}(2\mu_1)+e^{4i\Theta_k}\sqrt{\tilde{\xi}_1^2-4\mu_1^2}\big)  \\ 1 \end{pmatrix}
\end{aligned}
\end{equation}
Going back to (\ref{slowing swimming corr matrix eqn}) for the above single rotor type population and reducing the equation to the 2x2 system we get.
\begin{equation}  \label{slowing swimming corr matrix eqn 2}
\begin{aligned}
  L_{1,\bar{i}}X_{1,\bar{i}}+X_{1,\bar{i}}L_{1,\bar{i}}^H=-\hat{M}
\end{aligned}
\end{equation}
After eigen-decomposition with $L_{1,\bar{i}}=QDQ^{-1}$ we get: 
\begin{equation}  \label{slowing swimming corr matrix eqn 3}
\begin{aligned}
  D_{1,\bar{i}}Y+YD_{1,\bar{i}}^H=-Q^{-1}\hat{M}(Q^{-1})^{H}
\end{aligned}
\end{equation}
Where D is the diagonal eigenvalue matrix and $Y=Q^{-1}X(Q^{H})^{-1}$ where Q is the eigenvector matrix. 
\begin{equation} 
\begin{aligned}
\lambda_iY_{i,i}+\Bar{\lambda}_i{Y}_{i,i}=\begin{cases}
                                   \frac{\xi^2\hat{M}_{2,2}}{2(\tilde{\xi}_1^2-4\mu^2)}, & \text{if $\tilde{\xi}_1^2\geq 4\mu^2$} 
                                   \\
                                   -\frac{[4\mu^2+2i\mu\sqrt{\tilde{\xi}_1^2-4\mu^2}]}{2[\tilde{\xi}_1^2-4\mu^2]}\hat{M}_{2,2} & \text{if $\tilde{\xi}_1^2< 4\mu^2$}, \\
                                   \end{cases}
                                    \hspace{0.4 cm} 
\end{aligned}
\end{equation}
\begin{equation} 
\begin{aligned}
\lambda_1Y_{1,2}+\Bar{\lambda}_{2}{Y}_{1,2}=\begin{cases}
                                   \frac{[4\mu^2+2i\mu\sqrt{\tilde{\xi}_1^2-4\mu^2}]}{2[\tilde{\xi}_1^2-4\mu^2]}\hat{M}_{2,2}, & \text{if $\tilde{\xi}_1^2\geq 4\mu^2$} 
                                   \\
                                   -\frac{\xi^2\hat{M}_{2,2}}{2(\tilde{\xi}_1^2-4\mu^2)}, & \text{if $\tilde{\xi}_1^2< 4\mu^2$} \\
                                   \end{cases}
\end{aligned}
\end{equation}
\begin{equation} 
\begin{aligned}
\lambda_2Y_{2,1}+\Bar{\lambda}_{1}{Y}_{2,1}=\begin{cases}
                                   \frac{[4\mu^2-2i\mu\sqrt{\tilde{\xi}_1^2-4\mu^2}]}{2[\tilde{\xi}_1^2-4\mu^2]}\hat{M}_{2,2}, & \text{if $\tilde{\xi}_1^2\geq 4\mu^2$} 
                                   \\
                                   -\frac{\xi^2\hat{M}_{2,2}}{2(\tilde{\xi}_1^2-4\mu^2)}, & \text{if $\tilde{\xi}_1^2< 4\mu^2$} \\
                                   \end{cases}
\end{aligned}
\end{equation}
then 
\begin{equation} 
\begin{aligned}
Y_{1,1}=\begin{cases}
                                   \frac{\xi^2\hat{M}_{2,2}}{4(\tilde{\xi}_1^2-4\mu^2)[\sqrt{\tilde{\xi}_1^2-4\mu^2}+k^2D_t+\tilde{\xi}_1]}, & \text{if $\tilde{\xi}_1^2\geq 4\mu^2$} 
                                   \\
                                   -\frac{4\mu^2+2i\mu\sqrt{\tilde{\xi}_1^2-4\mu^2}}{4(\tilde{\xi}_1^2-4\mu^2)[k^2D_t+\tilde{\xi}_1]}\hat{M}_{2,2}, & \text{if $\tilde{\xi}_1^2< 4\mu^2$} \\
                                   \end{cases}
\end{aligned}
\end{equation}
\begin{equation} 
\begin{aligned}
Y_{2,2}=\begin{cases}
                                   -\frac{\xi^2\hat{M}_{2,2}}{4(\tilde{\xi}_1^2-4\mu^2)[\sqrt{\tilde{\xi}_1^2-4\mu^2}-k^2D_t-\tilde{\xi}_1]}, & \text{if $\tilde{\xi}_1^2\geq 4\mu^2$} 
                                   \\
                                   -\frac{4\mu^2-2i\mu\sqrt{\tilde{\xi}_1^2-4\mu^2}}{4(\tilde{\xi}_1^2-4\mu^2)[k^2D_t+\tilde{\xi}_1]}\hat{M}_{2,2}, & \text{if $\tilde{\xi}_1^2< 4\mu^2$} \\
                                   \end{cases}
\end{aligned}
\end{equation}                                   
with $\hat{M}_{2,2}=[\frac{2}{N_p}k^2D_t+\frac{8}{N_p}]$

For the non-diagonal terms:
\begin{equation} 
\begin{aligned}
Y_{1,2}=\begin{cases}
                                   \frac{[4\mu^2+2i\mu\sqrt{\tilde{\xi}_1^2-4\mu^2}]\hat{M}_{2,2}}{4[\tilde{\xi}_1^2-4\mu^2][-\tilde{\xi}_1-4-k^2D_t-\sqrt{\tilde{\xi}_1^2-4\mu_1^2}]}, & \text{if $\tilde{\xi}_1^2\geq 4\mu^2$} 
                                   \\
                                   -\frac{\xi^2\hat{M}_{2,2}}{4(\tilde{\xi}_1^2-4\mu^2)[-\tilde{\xi}_1-4-k^2D_t-\sqrt{\tilde{\xi}_1^2-4\mu_1^2}]}, & \text{if $\tilde{\xi}_1^2< 4\mu^2$} \\
                                   \end{cases}
\end{aligned}
\end{equation} 
\begin{equation} 
\begin{aligned}
Y_{2,1}=\begin{cases}
                                   \frac{[4\mu^2-2i\mu\sqrt{\tilde{\xi}_1^2-4\mu^2}]\hat{M}_{2,2}}{4[\tilde{\xi}_1^2-4\mu^2][-\tilde{\xi}_1-4D_r-k^2D_t+\sqrt{\tilde{\xi}_1^2-4\mu_1^2}]}, & \text{if $\tilde{\xi}_1^2\geq 4\mu^2$} 
                                   \\
                                   -\frac{\xi^2\hat{M}_{2,2}}{4(\tilde{\xi}_1^2-4\mu^2)[-\tilde{\xi}_1-4-k^2D_t+\sqrt{\tilde{\xi}_1^2-4\mu_1^2}]}, & \text{if $\tilde{\xi}_1^2< 4\mu^2$} \\
                                   \end{cases}
\end{aligned}
\end{equation} 
Note that $\tilde{\xi}_1$ is negative because the swimmers are pushers. We now are able to get $X=QYQ^H$ and although the expression for $X$ is involved we can gather from it analyical conclusions on spatial correlations for this parameter regime. 
\providecommand{\noopsort}[1]{}\providecommand{\singleletter}[1]{#1}%

\end{document}